\documentclass[final,5p,times,twocolumn]{elsarticle}

\usepackage{lineno,hyperref}
\usepackage{amsmath}
\usepackage{amssymb}
\usepackage{graphicx,xcolor}
\usepackage{enumitem}

\usepackage{lineno,hyperref}
\usepackage{amsthm}
\usepackage{xcolor}

\journal{Journal of Fusion Engineering and Design}

\begin{document}
\begin{frontmatter}
\title{Design and manufacturing of an optimized retro reflective marker for photogrammetric pose estimation in ITER}
\author[mymainaddress]{Laura Gon\c{c}alves Ribeiro \corref{mycorrespondingauthor}}
\cortext[mycorrespondingauthor]{Corresponding author}
\ead{laura.goncalvesribeiro@tuni.fi}
\author[mymainaddress]{Olli J. Suominen}
\author[thirdaddress]{Philip Bates}
\author[mymainaddress]{Sari Peltonen}
\author[mysecondaryaddress]{Emilio Ruiz Morales}
\author[mymainaddress]{Atanas Gotchev}
\address[mymainaddress]{Faculty of Information Technology and Communication, Tampere University, 33720 Tampere, Finland}
\address[mysecondaryaddress]{Fusion for Energy (F4E), ITER Delivery Department, Remote
Handling Project 
Team, 08019 Barcelona, Spain}

\address[thirdaddress]{Fusion for Energy, c/ Josep Pla, n°2 - Torres Diagonal Litoral - Edificio B3, 08019, Barcelona, Spain}

\begin{abstract}
Retro reflective markers can remarkably aid photogrammetry tasks in challenging visual environments. They have been demonstrated to be key enablers of pose estimation for remote handling in ITER. However, the strict requirements of the ITER environment have previously markedly constrained the design of such elements and limited their performance.
In this work, we identify several retro reflector designs based on the cat's eye principle that are applicable to the ITER usecase and propose a methodology for optimizing their performance. We circumvent some of the environmental constraints by changing the curvature radius and distance to the reflective surface.  We model, manufacture and test a marker that fulfils all the application requirements while achieving a gain of around 100\% in performance over the previous solution in the targeted working range. 
\end{abstract}


\end{frontmatter}
\section{Introduction}
A retro reflector is an optical element that returns incoming light in a cone around its source for a wide range of incident directions. Retro reflective (RR) markers have often been used in photogrammetry to provide reliable, high contrasting features in visually challenging environments~\cite{liu2012photogrammetric, clarke1994analysis}.
They are mainly characterised by their \textit{brilliancy} (reflectance for a specific observation angle), \textit{divergence} (maximum angular distribution of retro reflected light) and \textit{angularity} (brilliancy as a function of the incident angle).
An ideal retro reflector has the lowest divergence and highest angularity values. However, in photogrammetry, the light source and the sensor are usually not co-located. Therefore, a good retro reflector for such applications should have a divergence value that covers the viewing angles, without overly exceeding them.

The use of RR markers for pose estimation in ITER has been explored in a previous study~\cite{ribeiro2020robust}. The results of that work demonstrated that RR marker-enabled tracking simplifies the 3D node system~\cite{niu2019stereoscopic} by using a single camera only and achieves good performance in terms of accuracy, precision, and robustness.
The main objective of the previous study was to engineer and manufacture an RR marker from materials that can withstand the extreme environmental conditions, namely high temperature and radiation, the knuckle of the divertor's cassette locking system is exposed to. The proposed approach consisted of an array of spherical fused silica glass beads with a rear face coating, held by a stainless steel housing and a sieve-shaped cover. Each glass bead acts as a cat's eye retro reflector, where the primary lens focuses light into a reflective back layer.

\begin{figure}
\includegraphics[width=9cm]{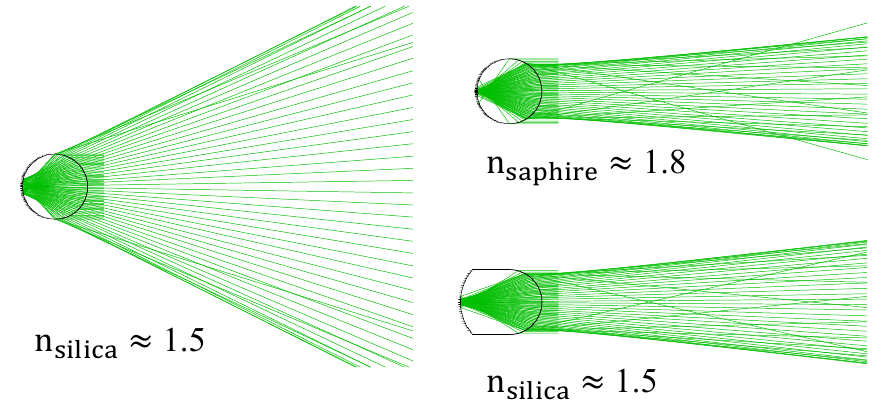}
\caption{Retro reflected light distribution of a fused silica glass bead (left side), narrower distribution produced by a type of glass with higher refractive index (upper right corner) and by manipulating the distance between the lens and the reflective surface.}
\label{fig:concept}
\end{figure}

The main problem with the earlier approach is that the proposed retro reflector model has only two degrees of freedom: the radius of the sphere (r) and the refractive index of the sphere’s material (n). 
Given the refractive index of fused silica ($n_{SiO2} = 1.4585$), the marker has a divergence of approximately 25 degrees, as illustrated in Figure~\ref{fig:concept}, left side, while the observation angles in the application can be led to be a few degrees, even at shorter working distances. 

The divergence could be optimised by using a type of glass with a higher refractive index, as shown in Figure~\ref{fig:concept}, upper right corner. However, only a very limited subset of optical glasses are suitable for the high total integrated gamma radiation doses ($>10MGy$) in the ITER divertor region. Most optical glasses rapidly become absorbing. Fused silica is overall the best option due to availability, cost, radiation resistance and the fact that it has already been tested and cleared for use in ITER. While using fused silica, it is not possible to optimize brilliancy at the viewing angles within the spherical glass bead design.

In this study, we aim at designing an alternative RR marker with narrower light distribution, which fulfills all the constraints of the application.
We hypothesize that this can be achieved having a model with higher number of degrees of freedom, allowing to better optimize brilliancy at the observation angles. More specifically, we focus on retro reflector designs were the distance between the focusing element and the reflective back layer can be manipulated, as shown in Figure~\ref{fig:concept}, lower right corner.

In Section~\ref{section:Methodology}, we present three marker types and propose a methodology for optimizing their performance. We analyse the pros and cons of each type and propose a manufacturing methodology for the most advantageous option.
In Section~\ref{section:Results}, we compare the performance of the developed solutions versus the state of the art using both ray tracing simulations and real data. 

\section{Methodology}
\label{section:Methodology}
\subsection{Retro reflective marker designs}

\begin{figure}
\includegraphics[width=9cm]{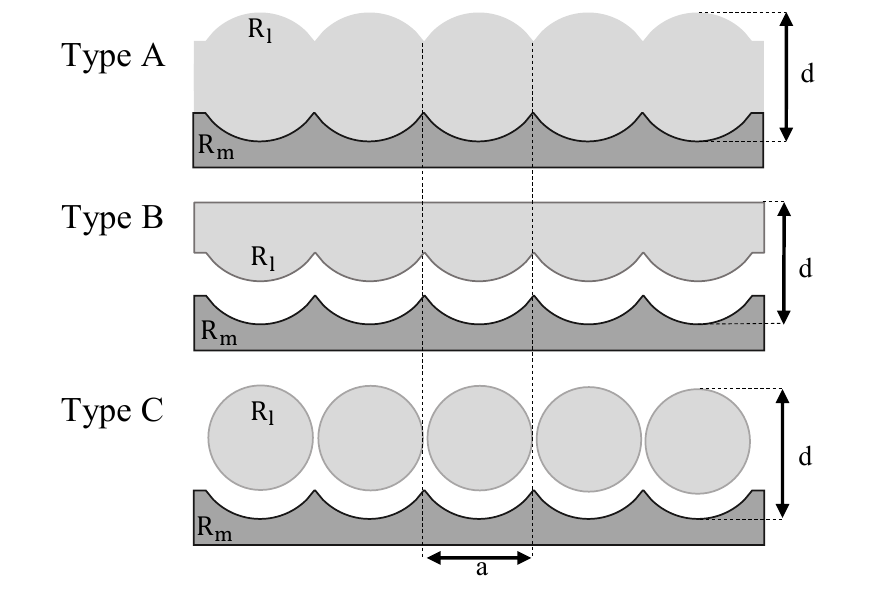}
\caption{Retro reflector design alternatives: based on a biconvex micro lens array (type A), a plano convex microlens array (type B), and an array of ball lenses (type C).}
\label{fig:types_RR}
\end{figure}

The retro reflector design proposed in~\cite{lundvall2003high} is used as a starting point in our work. 
It is composed of a front layer of densely packed convex spherical micro lenses, a back surface of densely packed concave spherical micro mirrors and a transparent layer in between. 
The lens and the mirror have coinciding optical axes, a common aperture diameter and each lens and corresponding mirror form one retro reflecting element.
The distance between the primary lens and secondary spherical mirror can be adjusted.
This design is essentially based on biconvex microlens. We refer to it as "type A" in Figure~\ref{fig:types_RR}. It five degrees of freedom as follows:

\begin{itemize}[noitemsep]
	\item the curvature radius of the focusing lens ($R_l$),
	\item the curvature radius of the mirror ($R_m$),
	\item the aperture diameter of the lens and mirror (a),
	\item and the refractive index of the spacer material (n).
\end{itemize}

The main challenge we anticipate with this approach is that the custom manufacturing of biconvex fused silica micro lenses is rather resource consuming. The tooling costs at the prototyping stages are high for any manufacturing techniques we are aware of. 
Further, the higher volume production is sill expensive for the relatively small amount of units that the application requires.
As a solution to these impediments, we consider two additional marker designs where it is still possible to vary the distance, d, and the curvature radius of the focusing and reflective surfaces, $R_l$ and $R_m$.

\begin{figure}
\includegraphics[width=9cm]{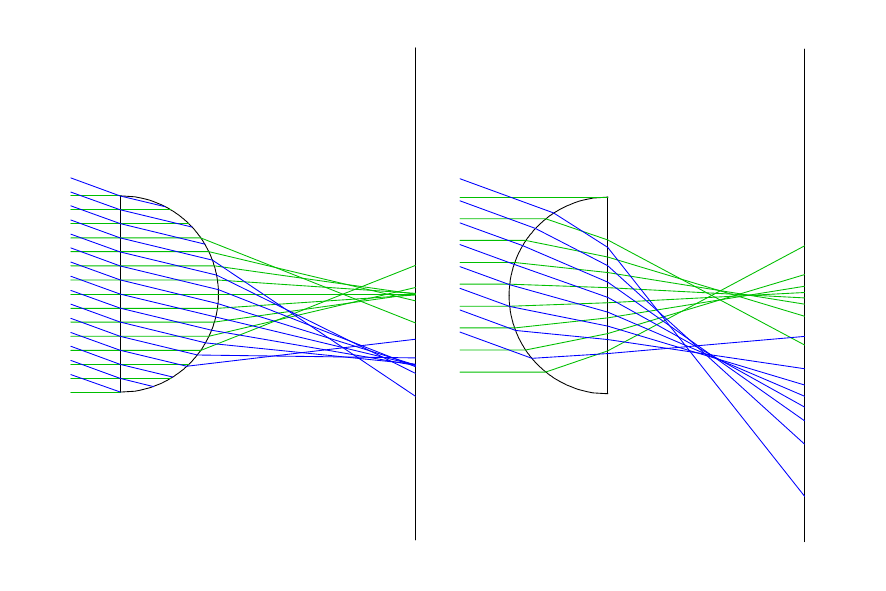}
\caption{Refration of incoming light rays of different orientations by a plano convex (left) and convex plano (right) lens.}
\label{fig:subtypesB}
\end{figure}

Design type B considers the use of a plano convex micro lens array to focus light onto a reflective piece. We expect that an off-the-shelf specimen can potentially be used, which would reduce the cost of the prototyping and manufacturing stages. 
The convex planar option, where the micro lens array is flipped by $180^{\circ}$, would have some desirable characteristics in relation to its counterpart, such as lower comatic and spherical aberration. However, this option suffers from a high value of astigmatism and was discarded for this reason. We see the effects of astigmatism in Figure~\ref{fig:subtypesB} (right side), where the focus points fall onto a parabolic surface and do not satisfy the Petzval condition~\cite{hecht2002optics}.

Design type C considers the use of several spherical glass lenses that are aligned and kept at a given distance to a reflective plate. The main advantage of this approach is the ease of manufacturing and consequent low cost of spherical glass beads and their intrinsically high aperture.
The main disadvantage is the increased challenge in manufacturing a housing where each individual lens is held in place and aligned to the corresponding reflective surface. The alignment requires particularly high precision if the ball lenses have small dimensions.

The use of conventionally available reflective coatings (e.g. gold and silver) in the ITER environment adds risk, both due radiation resistance of the coating itself, or the extra materials needed to adhere the coating to the glass surface, as due to concerns about activation of the adhering materials. Therefore, for all marker types, the reflective part shall consist of a polished metal piece that has been indented with the correct curvatures.

\subsection{Design constraints}
\label{subsection:design_constraints}

\begin{figure}
\includegraphics[width=9cm]{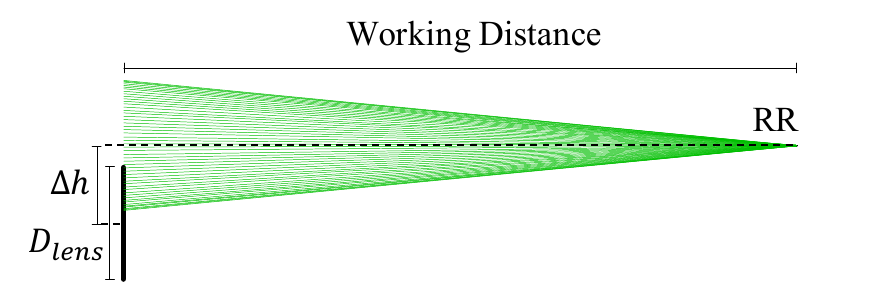}
\caption{Plan view of design constraints. We aim at optimizing the number of rays arriving at a camera lens with diameter $D_{lens}$, which is displaced from the light source by a fixed distance $ \Delta h$. The working distance is variable within a certain range.}
\label{fig:setup_simulations}
\end{figure}

While designing an optimized RR marker for photogrammetry, the objective is to maximize the percentage of incident light rays arriving at the camera lens. Therefore, as a starting step, the size of the camera lens, the distance of the camera to the lighting element and the working distance, as seen in Figure~\ref{fig:setup_simulations}, need to be specified. 

\begin{figure}
\includegraphics[width=9cm]{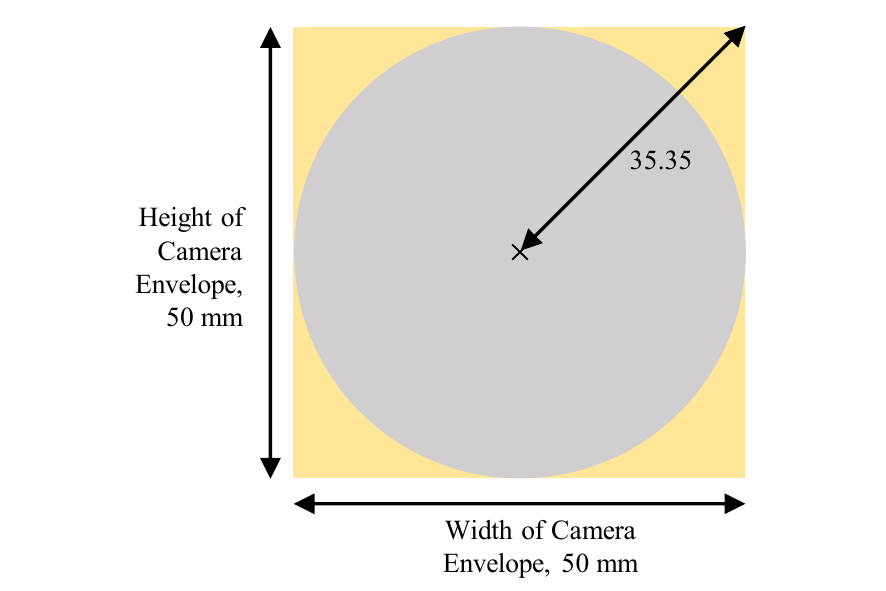}
\caption{Plan view of camera and lighting element envelope with maximum possible lens footprint.}
\label{fig:size_envelope}
\end{figure}

In our target use case, the camera is to be installed on the wrist of a robotic manipulator, where space is severely limited. The maximum allowable envelope for the camera and lighting is 50$\times$50 mm. Within this envelope we can have, at most, a lens radius of 25 mm and maximum distance between the lens and the lamp of approximately 35 mm, as shown in Figure~\ref{fig:size_envelope}.

Further, the working distance is fixed based on the available space, to a range of 300 to 500 mm.
As angularity goes, we aim at maximizing the uniformity of light return for entrance angles in the range of $-20^{\circ}$ to $20^{\circ}$.

\subsection{Optimization of retro reflector performance} 

The proposed methodology for designing an optimized cat's eye retro reflector consists of the following sequential steps:

\bigbreak

\textbf{Step 1}: Establishing the size of each individual retro reflecting unit.
The size of each individual retro reflecting unit determines what is the maximum possible aperture size. The maximum aperture size can be chosen based on the camera resolution and working distance, such that the individual structures that constitute one marker are indistinguishable in the camera view. This will lead to the marker being more easily perceived, at later stages, as a uniform object. 
However, there is a trade off between the size of the features and the manufacturing challenges.
The smaller the size of the features, the tighter the requirements on the manufacturing tolerances will be to ensure the designed functionality of the piece, particularly when it comes to the alignment of the optical axes of lens and mirror.
As the aperture can only be as large as the curvature radius of the lens, it follows that $R_l = a_{max}/2$.
In the conditions described in Subsection~\ref{subsection:design_constraints}, and considering a typical camera resolution of approximately 640$\times$480 pixels, we start with $a_{max}$ = 0.4 mm and $R_l$ = 0.2 mm.

For marker type C, this step corresponds to the selection of the size of the glass spheres. In this case, it is important to consider the additional manufacturing challenges in holding and aligning several independent elements of small dimensions at a fixed distance to the reflective plate. For this reason, the use of larger sized ball lenses might be worth considering, even if it leads to the final RR element not being perceived as fully uniform.

\bigbreak

\textbf{Step 2}: Establishing the distance between the lens and the reflective surface (d), based on the distance at which the paraxial rays converge ($d_f$).

The paraxial focal length of the lens can be calculated, for each design type, as described below~\cite{hecht2002optics}.

\begin{itemize}
	\item For type A, i.e. the biconvex lenses case, there is one interaction in one single interface, as shown in Figure~\ref{fig:generic_lens_biconvex}:

    \begin{equation}
    d_f=\frac{n R_l}{n-1}.
    \end{equation}
    
    \begin{figure}
    \includegraphics[width=9cm]{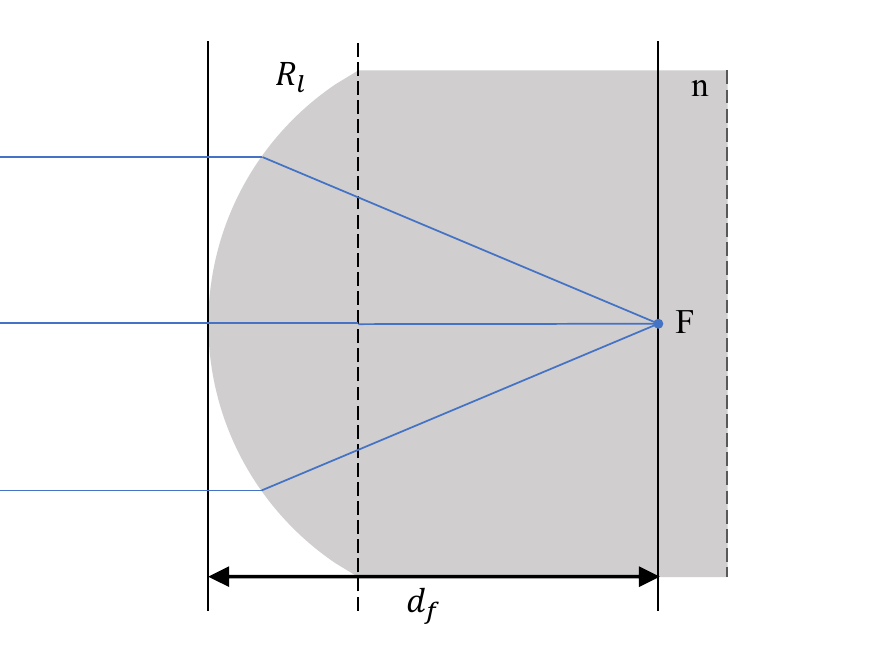}
    \caption{Generic lens structure composed of one curved medium interface with curvature radius $R_l$. The lens has refractive index n and the surrounding medium has refractive index $n_{air}$. Paraxial focus is at the lens medium, at distance $d_f$ from the first interface.}
    \label{fig:generic_lens_biconvex}
    \end{figure}
	
	\item For type B, i.e. the plano convex lenses case, the first interface is a plane ($r_1 = \infty$), as shown in Figure~\ref{fig:generic_lens}:
	
	The power of each individual interface ($K1$, $K2$) and the total power of the lens (K) are as follows:
	
	\begin{equation}
	K_1=\ 0,\ K_2=-\frac{n-1}{-R_l},\
	\end{equation}
	
	\begin{equation}
	K=K_1+K_2-K_1K_2\frac{d}{n},
	\end{equation}
	
	\begin{equation}
	K=\frac{n-1}{R_l}.
	\end{equation}
	
	The effective focal length is given by:
	\begin{equation}
	f=\frac{1}{K}=\frac{R_l}{n-1}
	\end{equation}
	
	and the back focal length (BFL) is given by:
	
	\begin{equation}
	BFL=\ f\frac{1}{n}\ = \frac{R_l}{n(n-1)}.
	\end{equation}
	
	Therefore, 
	
	\begin{equation}
	d_f= BFL + d = \frac{R_l}{n(n-1)} + d.
	\end{equation}
	
	\begin{figure}
    \includegraphics[width=9cm]{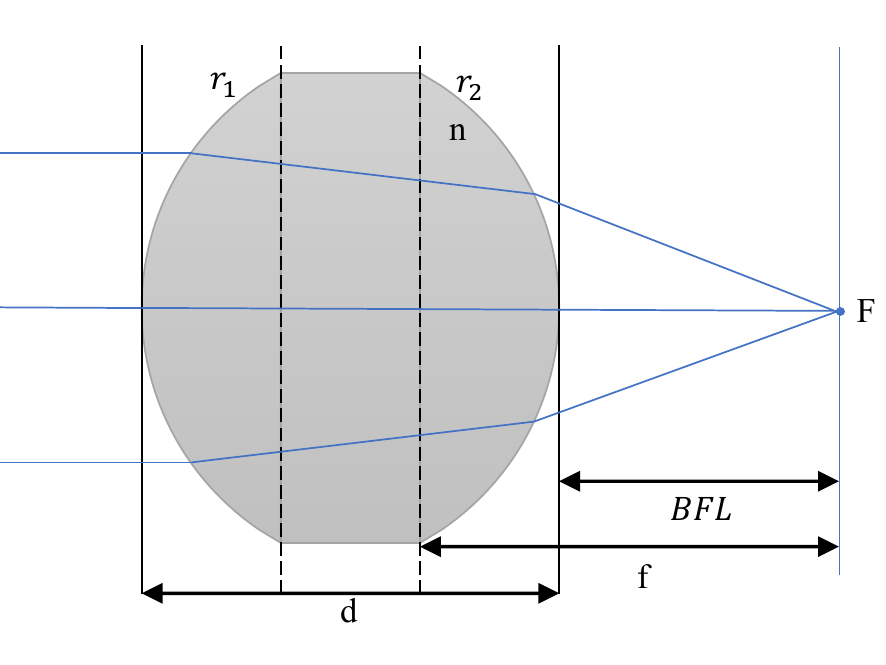}
    \caption{Generic lens structure composed of two curved medium interfaces with radii $r_1$ and $r_2$ and a total thickness d. The lens has refractive index n and the surrounding medium has refractive index $n_{air}$. Paraxial focus is at the outgoing medium, at distance BFL from the last interface.}
    \label{fig:generic_lens}
    \end{figure}
	
	\item For type C, i.e. the ball lenses case, $r_1 = r_2 = R_l$, as shown in Figure~\ref{fig:generic_lens}.
	
	The power of each individual interface (K1, K2) and the total power of the lens (K) are:
	
	\begin{equation}
	K_1=\ \frac{n-1}{R_l},\ K_2=-\frac{n-1}{-R_l},\ 
	\end{equation}
	
	\begin{equation} 
    \begin{split}
    K & = K_1+K_2-\ K_1K_2\frac{2R_l}{n} \\
     & = \frac{n-1}{R_l}+\frac{n-1}{R_l}-\left(\frac{n-1}{R_l}\right)^2\frac{2R_l}{n} \\
     & = \frac{2n^2-2n-2n^2+4n-2\ }{n R_l} \\
     & = \frac{2(n-1)}{n R_l}.
    \end{split}
    \end{equation}
    
    The effective focal length is given by:
	\begin{equation}
	f=\frac{1}{K}=\frac{n R_l}{2(n-1)}
	\end{equation}
	
	and the BFL is given by:
	
	\begin{equation}
	BFL= f \frac{2-n}{n}=\frac{R_l(2-n)}{2(n-1)}.
	\end{equation}
	
    Therefore, 
	
	\begin{equation}
	d_f= BFL + d = \frac{r(2-n)}{2(n-1)} + d.
	\end{equation}
	
\end{itemize}

\begin{figure}
\includegraphics[width=9cm]{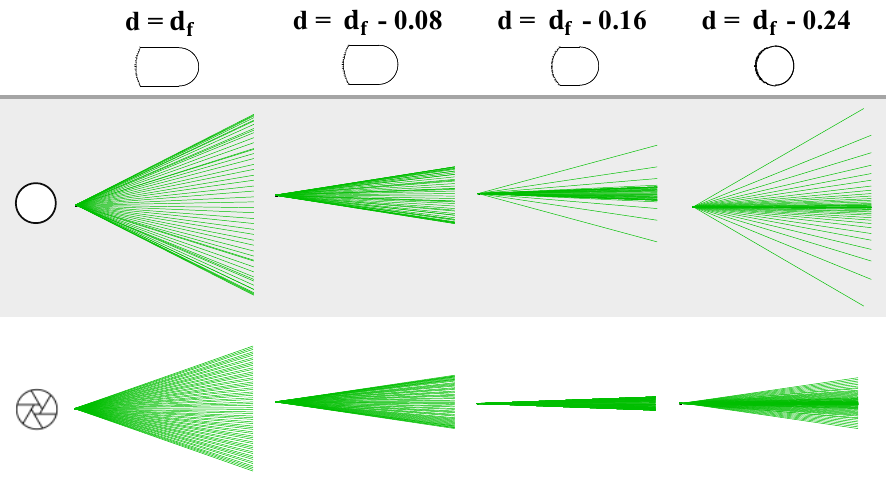}
\caption{Light ray distributions for different RR marker configurations of type A. The distance to the reflective surface (d) varies in relation to the focus distance ($d_f$). In the top row we show the maximum (0.4 mm), and on the bottom row a smaller aperture.}
\label{fig:lightpattern_biconvex}
\vspace*{1cm}
\includegraphics[width=9cm]{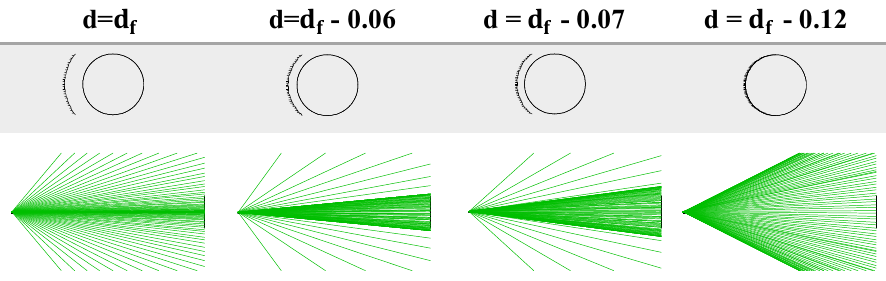}
\caption{Light ray distributions for different RR marker configurations of type C. The distance to the reflective surface (d) is varied in relation to the focus distance ($d_f$). The spheres have a radius of 0.2 mm.}
\label{fig:lightpattern_ball}
\end{figure}

The distance to the mirror (d) can be set equal or different to the focus distance ($d_f$), depending on the desired optical behaviour.
In Figure~\ref{fig:lightpattern_biconvex} we show the light ray distributions corresponding to different configurations ($d = d_f$, $d = d_f-0.08$, $d = d_f-0.16$ and $d = d_f-0.24$) within marker type A. Similar information is presented in Figure~\ref{fig:lightpattern_ball} for marker type C.
We find that specific configurations with $d < d_f$ produce a narrower cone of retro reflection than $d = d_f$. This allows us to tailor the divergence to the use case to a certain degree. Further, in these same configurations we observe a light ray distribution that favours the periphery of the cone. This type of distribution is advantageous in further optimizing the number of rays arriving at the lens.

\begin{figure}
\includegraphics[width=9cm]{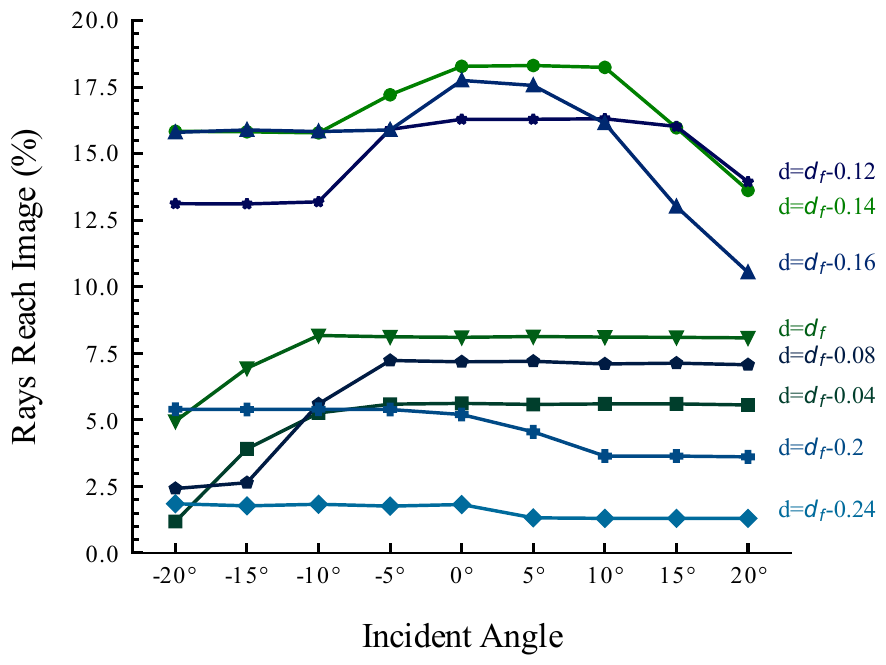}
\caption{Behaviour of RR marker configurations of type A with varying distance to the reflective surface. The curvature radius and aperture of the primary lens are constant with $R_l$ =0.2 and a=0.4. The curvature radius of the reflective surface, $R_m$, is changed accordingly, to fulfill the condition of retro reflection.}
\label{fig:biconvex_varryd}
\end{figure}

Figure~\ref{fig:biconvex_varryd} quantifies the percentage of incident rays arriving at the circular surface of the camera lens for several marker configurations within type A. The data in this and in subsequent figures in Section~\ref{section:Methodology}, is obtained using a commercial sequential raytracing software (OSLO by Lambdares~\cite{oslo}), considering a single light source that is vertically displaced in relation to the circular surface of the lens. The displacement distance is as described in Section~\ref{subsection:design_constraints}.

A similar analysis can be easily obtained for the other two configurations. From this type of representation the most advantageous configuration can be easily determined.
While Figure~\ref{fig:biconvex_varryd} shows the percentage of returned rays for a single working distance, in determining the most advantageous configuration the full range of working distances must be considered. 

\textbf{Step 3}: Fulfilling the retro refection condition. 
Due to Petzval field curvature aberration, as we go further away from the optical axis, the focal point does not actually fall on a focal plane at distance $d_f$, but rather on a spherical surface. In order to obtain a retro reflector that keeps its RR behaviour over a wide range of incident directions, the reflective surface should accompany this curvature~\cite{blechinger2005handbook}. This ensures that focus is kept at the same distance relative to the reflective surface regardless of the incident angle.

In the case where the reflective surface is placed at the focus distance, its curvature radius should equal the Petzval radius of the lens.
For a generic $d$, the condition of retro reflection can be defined as a function of the Petzval radius ($R_{Petzval}$) and the distances $d$ and $d_f$, for each marker type, as described below.

\begin{itemize}
	\item For type A, the condition of retro reflection is as follows:

    \begin{equation}
    R_m=d-R_l.
    \end{equation}
	
	\item For type B, the Petzval radius is:
	
	\begin{equation}
	R_{Petzval}=- \frac{n\ R_l}{n-1}
	\end{equation}
    
    and the retro reflection condition is: 
    
    \begin{equation}
    \begin{split}
    R_m & = - \frac{n\ R_l}{n-1} -BFL+d \\
     & = - \frac{n\ R_l}{n-1} - \frac{R_l}{n(n-1)} + d \\
     & = - \frac{R_l(n+1)}{n} + d.
    \end{split}
    \end{equation}
	
	\item For type C, the Petzval radius is:
	
	\begin{equation}
	R_{Petzval}=\frac{nR}{2\left(n-1\right)}
	\end{equation}
	
	and the retro reflection condition: 
    
    \begin{equation}
    \begin{split}
    R_m & = \frac{nR}{2\left(n-1\right)} -BFL+d \\
     & = \frac{nR}{2\left(n-1\right)} - \frac{r(2-n)}{2(n-1)} + d \\
     & = R_l + d.
    \end{split}
    \end{equation}
	\end{itemize}

\bigbreak

\textbf{Step 4}: Establishing the final aperture value ($a$) within the range $0<a<a_{max}$. 

\begin{figure}
\includegraphics[width=9cm]{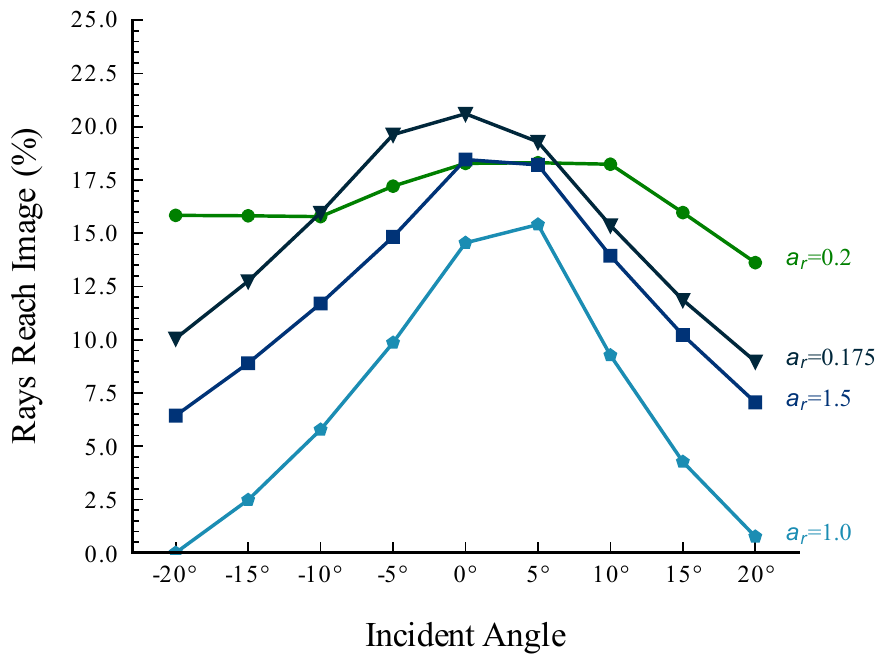}
\caption{Behaviour of RR marker configurations of type A with varying aperture radius ($a_r$). The curvature radius of the primary lens and reflective surface, and the distance between them are constant with $R_l$ =0.2, d = $d_f$-0.14 and $R_m$ = 0.3.}
\label{fig:biconvex_varrya}
\end{figure}

There is an evident trade-off in the establishment of the aperture value. On one hand, smaller aperture values lead to lower divergence through the minimization of the effects of spherical aberration. On the other hand, smaller aperture values are associated with worse angularity. This trade-off is evident in Figure~\ref{fig:biconvex_varrya}, which quantifies the percentage of incident light rays arriving at the lens surface for several marker configurations of type A.

For this reason, we opt to manipulate the divergence of the RR by changing the distance to the reflective surface (in step 2) and tend to keep the aperture at the highest possible value. This allows us to achieve simultaneously lower divergence and higher angularity values, leading to a higher return in the desired range of incident angles.

In an ideal case, the maximum entrance angle occurs when the paraxial focus strikes the edge of the reflective surface. However, when considering the effects of spherical aberration the outcome becomes less favourable. There may be a considerable percentage of reflected rays that fall out of the aperture of the primary lens, degrading the angular performance. For this reason we consider that the type of analysis in Figure~\ref{fig:biconvex_varrya} is more informative than the calculation of the maximum entrance angle.

\subsection{Choosing the most advantageous design type}
\begin{figure}
\includegraphics[width=9cm]{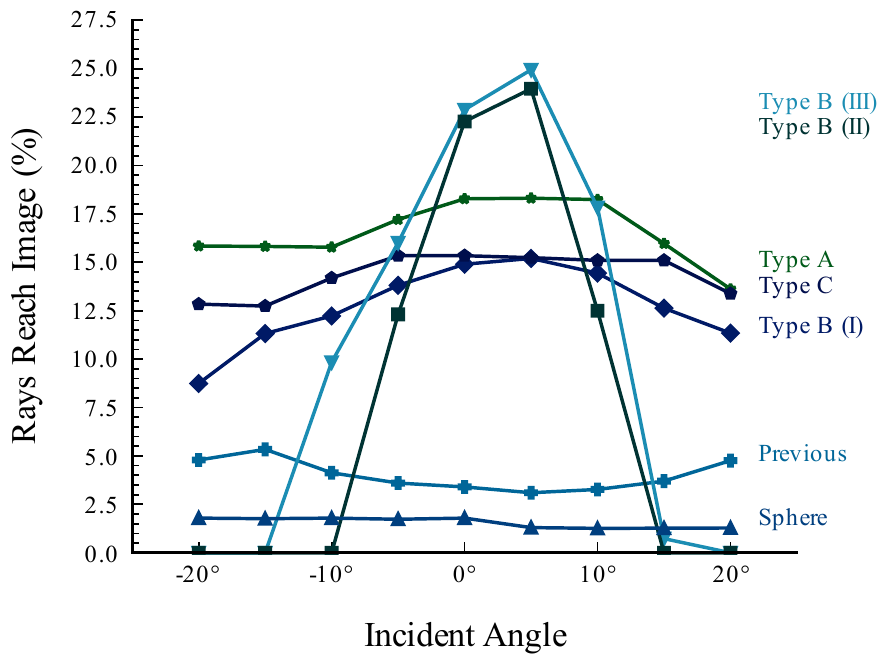}
\caption{Behaviour of selected marker configurations of each type. Type A with $R_l=0.2$ and $d = d_f - 0.14$, type B (I) with $R_l = 0.2$ and $d = d_f - 0.03$, and type C with $R_l = 0.5$ and $d = d_f - 0.15$. Type B (II) and Type B (III) correspond to the two off the shelf varieties. For these types $R_m$ is changed accordingly to fulfill the retro reflection conditions. "Previous" represents the previous solution, where $R_l = 1$ and the reflective plate is flat ($R_m =0$). "Sphere" represents the traditional cat's eye retro reflector where the reflective plate follows the curvature of the sphere with $R_l = R_m = 1$.}
\label{fig:compare_all}
\end{figure}

As seen in Figure~\ref{fig:compare_all}, we are often able to optimize the retro reflectivity values of type A better than those of types B and C.

\begin{figure}
\includegraphics[width=9cm]{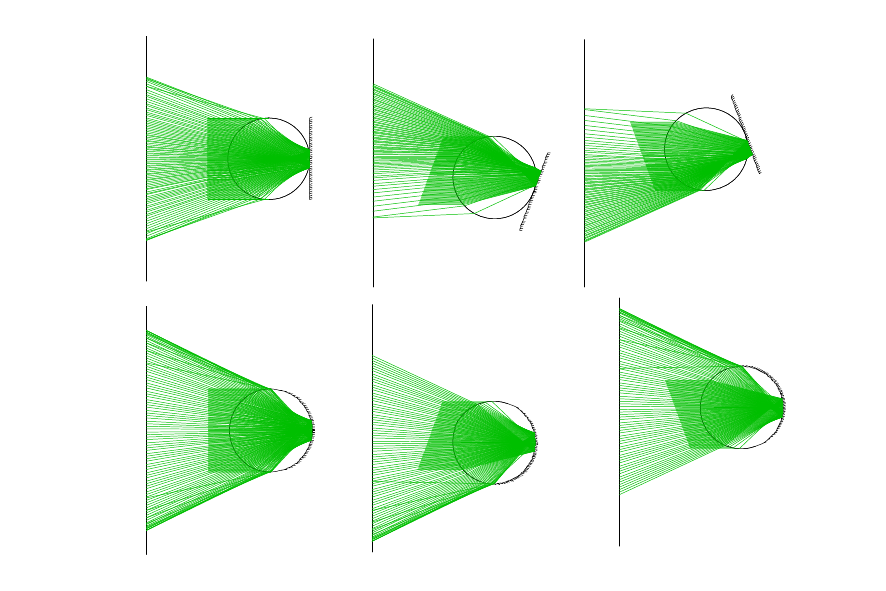}
\caption{Light pattern created by the previous marker, where the reflective plate is flat (top), and by a standard cat's eye retro reflector, where the curvature of the reflective surface accompanies the curvature of the sphere (bottom).}
\label{fig:coatedVSflat}
\end{figure}

However, marker types B and C do not fall far behind and have the potential to significantly improve upon the previous approach. 
The previous approach differs from a standard cat's eye retro reflector in the backing plate being flat, rather than accompanying the curvature of the sphere. This creates an asymmetrical retro reflected light distribution, as demonstrated in Figure~\ref{fig:coatedVSflat}. This distribution slightly improves performance in our application in relation to the case where the backing accompanies the curvature of the sphere.

We found that it is uncommon to find off-the-shelf varieties of plano convex micro lens arrays with very high effective lens aperture values in comparison to their radius of curvature (also referred to as high lens sag values). In Table ~\ref{table:1}, we describe the characteristics of the two off-the-shelf lens arrays we found with acceptable dimensions. These types are optimized following steps 2 to 3 in our proposed methodology.

\begin{table}
\begin{center}
\caption{Radius of curvature of the lens ($R_l$), aperture (a) and thickness (t) of two off-the-shelf plano convex micro lens varieties.}
\begin{tabular}{c|c|c|c} 
 & $R_l$ & $a$ & $t$ \\
\hline
II (Okotech) & 0.425 & 0.111 & 0.05 \\
\hline
III (Suss Microoptics) & 0.35 & 0.25 & 0.9 \\
\hline
\end{tabular}
\label{table:1}
\end{center}
\end{table}

We see in Figure~\ref{fig:compare_all} that of these two varieties, III has, as expected, better angular performance. While the plano convex options have a more limited angular performance than the remaining options, they reach higher values of retro reflectance at smaller viewing angles.

However, there is an additional challenge with this approach that we have not yet considered. Often, for cost efficiency, these arrays are cut from larger sheets and the distribution of lenses in the array is not exactly known. This constitutes a problem since the exact distance between each lens and the edge of the array needs to be known to construct the backing plate of corresponding features. Measuring the distance of the first lens to the edge of the array would be possible, although not trivial due to the small size of the lenses. Further, it would be highly inefficient to measure, model and manufacture a marker housing for each individual lens array piece. This led us to discard the use of an off the shelf plano convex lens array from our subsequent studies.

\begin{figure}
\includegraphics[width=9cm]{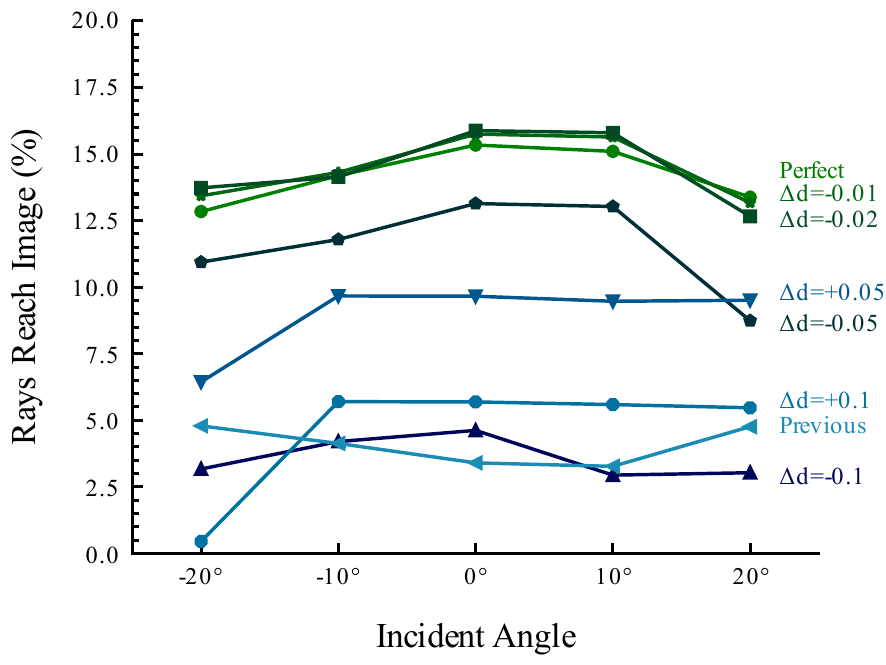}
\caption{Robustness of selected marker configuration of type C ($R_l = 0.5$, $d = d_f - 0.15$, $R_m = 0.65$) to errors in the distance to the reflective surface ($\Delta d$).}
\label{fig:deltad}
\end{figure}

For our intended use case, it seems that marker type C would have the potential for the best cost-performance trade-off and, moving forward, we focus on this approach.  
Figure~\ref{fig:deltad} shows the robustness of the selected type C. It instructs that the proposed type has considerable advantages over the previous solution. One can also see that the design is quite tolerant to errors in the distance to the reflective plate, d. In fact, even for very significant errors ($\Delta$d = 0.1 mm for d = 0.15 mm), the behaviour is not worse than the previous solution.
When it comes to the alignment of the optical axes of lenses and reflective surfaces, we consider that deviations in the order of 10 $\mu m$ cause no significant changes to the behaviour of the retro reflector.

\subsection{Manufacturing process}
\begin{figure}
\begin{center}
\includegraphics[width=9cm]{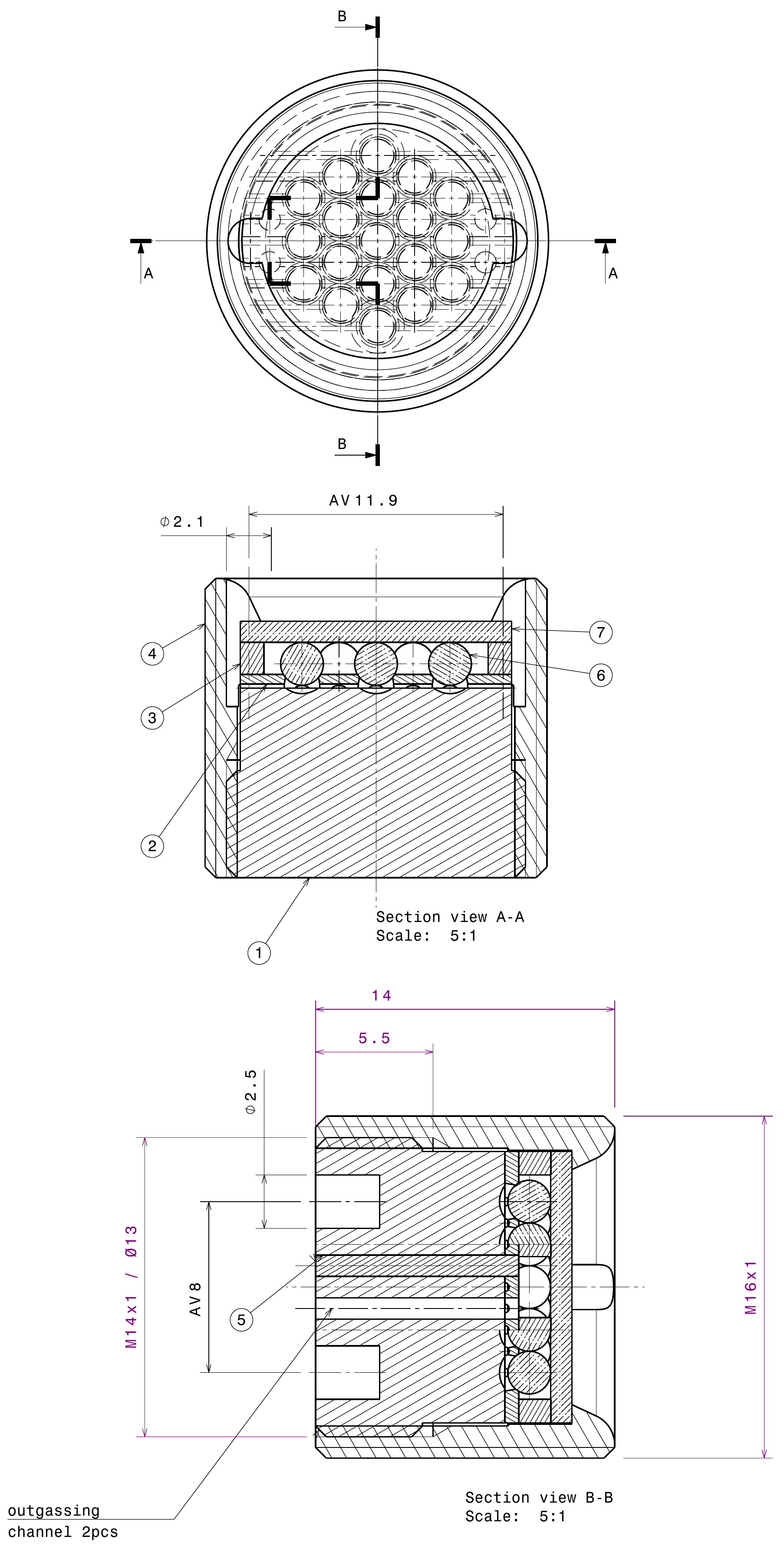}
\caption{Diagram of full marker assembly. 1- Reflective Base. 2-Spacer Disk. 3- Spacer Ring. 4- Body. 5- Guiding Pins. 6-Spherical Glass Beads. 7-Glass Window.}
\label{fig:fullassembly}
\end{center}
\end{figure}

In this section we describe the manufacturing process of the retro reflective marker housing, the reflective backing and the required holding structures (Figure~\ref{fig:fullassembly}). 

\begin{figure}
\begin{center}
\includegraphics[width=6cm]{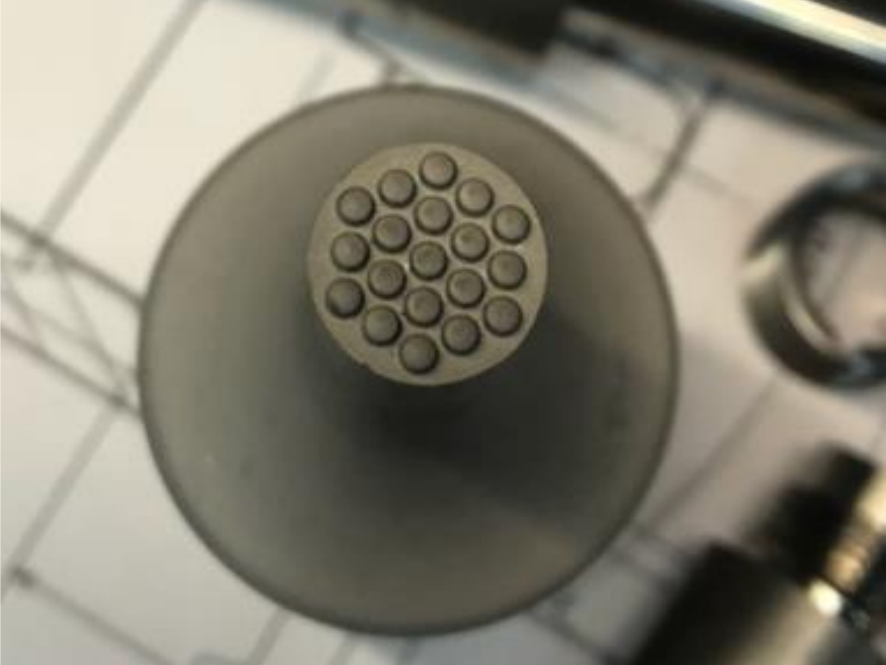}
\caption{Pressing tool used to indent the array of spherical reflective surfaces into the stainless steel backing.}
\label{fig:tool}
\end{center}
\end{figure}

A tool with inverse features of the reflective backing array (Figure~\ref{fig:tool}) has been produced by CNC machining and is applied through hydraulic pressing to indent the reflector array into a stainless-steel block (Figure~\ref{fig:fullassembly} - part 1). Four holes are added by EDM wire cutting on the same block – two holes for the guiding pins used to align the assembly (Figure~\ref{fig:fullassembly} - part 5), and two holes to act as outgassing channels. Outgassing channels are needed due to the vacuum conditions in ITER. The resulting block is milled to size and threaded on the outer surface.

The distance between the reflective surface and the reflective beads is ensured by a spacer, i.e. a plate with suitably sized holes matching the reflector pattern (Figure~\ref{fig:fullassembly} - part 2). This plate is made by EDM wire cutting. Outgassing channels and guiding pin holes are added using the same method. To ensure repeatability, a cylinder with the through holes is first made and then sliced to the specified thickness to create the spacer discs. 
Considering the expected thermal expansion of materials in the marker (fused silica has a lower thermal expansion coefficient than stainless steel) and to prevent damage of the glass beads resting against sharp edges, the holes' inner edges are bevelled to match the curvature of the beads.

The marker is encapsulated in a stainless-steel cylinder (Figure~\ref{fig:fullassembly} - part 4) milled to be hollow and threaded with the corresponding thread to fit the reflector piece. The spacer disc rests on top of the reflector held in place by guiding pins cut from 1 mm thick stainless steel welding wire, and the glass beads (Figure~\ref{fig:fullassembly} - part 6) rest on their respective holes on the spacer. 
In order to hold the glass spheres without using a sieve-like structure that would reduce the reflective area, a glass window is introduced in the design of the housing (Figure~\ref{fig:fullassembly} - part 7).
The topmost feature is a shoulder to keep in place the glass window. An EDM cut spacer ring (Figure~\ref{fig:fullassembly} - part 3) is added to avoid pressure of the glass beads on the glass window.


\section{Results}
\label{section:Results}
\subsection{Raytracing simulations}
\begin{figure}
\includegraphics[width=9cm]{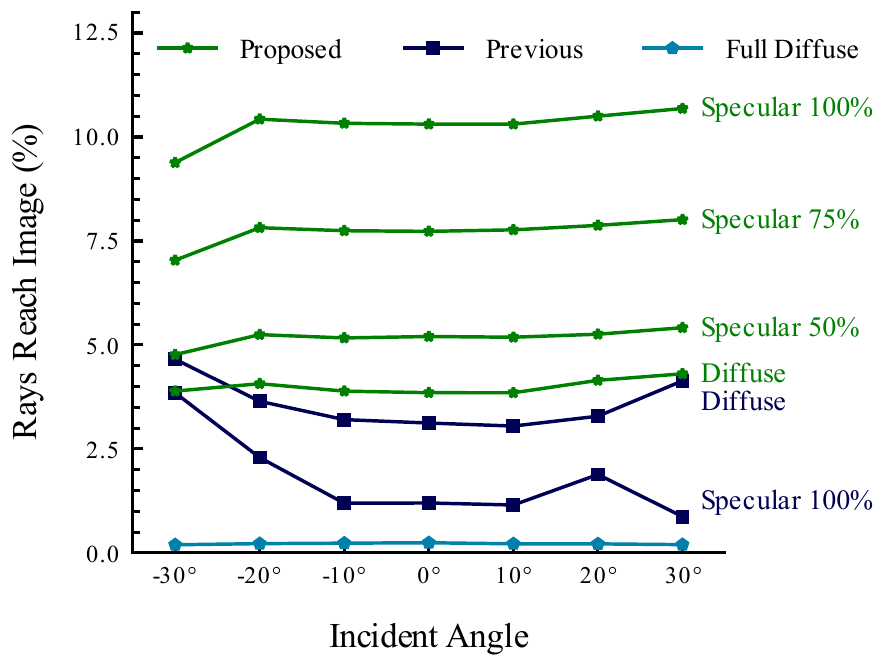}
\caption{Behaviour of the previous and proposed markers for different optical characteristics of their respective reflective surfaces according to ray-tracing simulations. "Diffuse" refers to a fully diffuse, non RR marker.}
\label{fig:results_tracepro}
\end{figure}

The analysis done thus far assumes a single perfect specular reflection at the reflective surface of the marker. This is a simplification of the problem at hand. Light can, in fact, get scattered and reflected multiple times at different interfaces, the reflection might not be perfectly specular, and some light might be absorbed by the materials.
To get a more realistic understanding of the comparative performance of the RR markers, we analysed the performance of the previous and the newly developed marker using a commercial Monte Carlo ray tracing software (TracePro by Lambdares~\cite{tracepro}). 
We demonstrate the effect of different optical characteristics of the marker's backing plates in Figure~\ref{fig:results_tracepro}.
“Full Diffuse” refers to the behaviour of a non retro reflective, diffuse, white element without any optical elements in front.  

\subsection{Experimental evaluation}

\begin{figure}
\includegraphics[width=9cm]{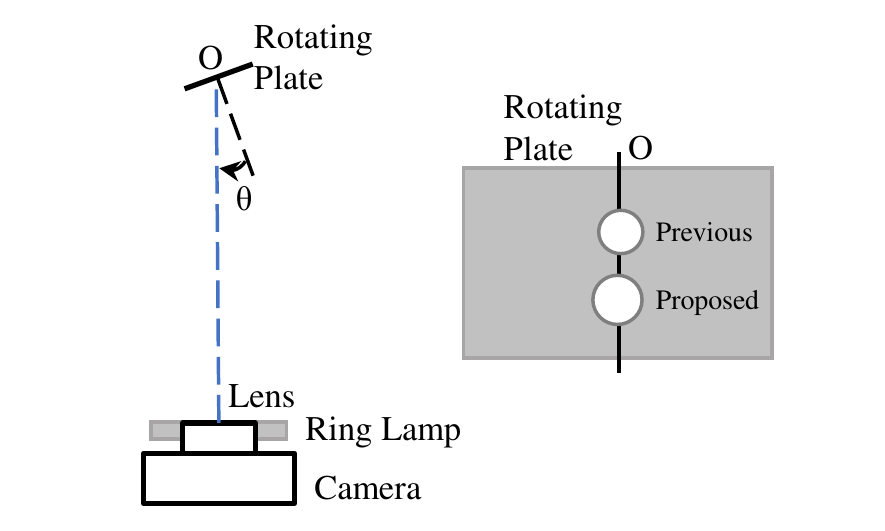}
\caption{Experimental setup for RR marker testing (not to scale). Samples are rotated around centre point O by angle $\theta$. Camera and LED sources are stationary throughout the experiments.}
\label{fig:experimental_setup}
\end{figure}

The angular response of RR markers is measured using a digital camera and a ring light. With this test setup we obtain response curves that are meaningful and directly applicable to photogrammetry use cases. The response is evaluated for increasing entrance and viewing angles by rotating the test samples around their centre points. The experimental setup is depicted in Figure~\ref{fig:experimental_setup}.

The camera has 1920$\times$1200 resolution and the lens has a horizontal field of view of 71.2 degrees. The camera resolution corresponds to approximately 0.4 mm per pixel at the sampling distance of 300 mm and 0.5 mm per pixel at the sampling distance of 500 mm.

\begin{figure}
\begin{center}
\includegraphics[width=7cm]{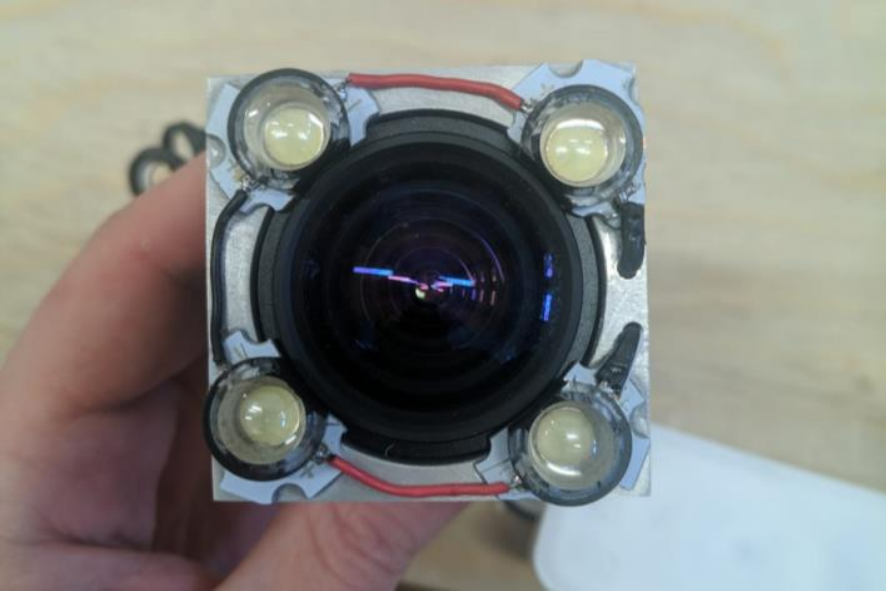}
\caption{Light setup around camera lens composed of four LEDs at the corners of the camera envelope. Distance from the centre of each LED to the centre of the camera lens is 22 mm. Diameter of the camera lens is 30 mm.}
\label{fig:lamp_picture}
\end{center}
\end{figure}

We compute the response curve at two discrete values (300 mm and 500 mm), corresponding to the limits of the targeted working range. 

The setup of light sources is as seen in Figure~\ref{fig:lamp_picture}, with two sets of two diametrically opposite LEDs. The distance from the centre of each LED to the centre of the camera lens is 22 mm. The observation angle (between the centre of the camera and the light source) is, therefore, fixed for a given working distance and is approximately 4 degrees at 300 mm and 3 degrees at 500 mm.
Response curves are symmetrical around the 0 degree incident angle, due to the added contributions of opposing light sources.

In designing the RR marker, the optimization procedure described in the methodology section was followed for these conditions.

\begin{figure}
\includegraphics[width=9cm]{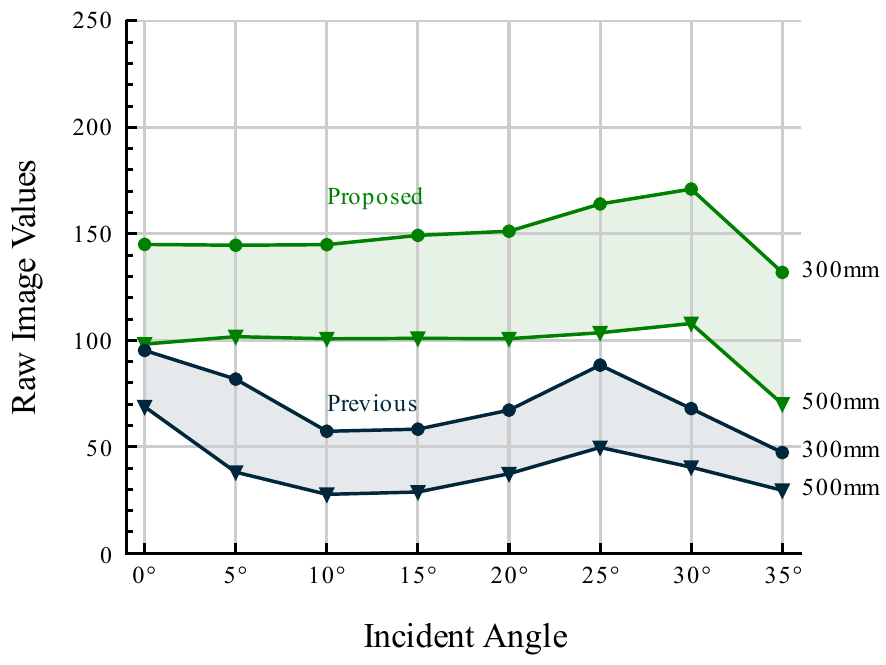}
\caption{Experimental evaluation of average intensity of all pixels belonging to the marker, in raw image values.}
\label{fig:experimental_results}
\end{figure}

The response curves in Figure~\ref{fig:experimental_results} are in line with what was anticipated in the theoretical analysis. One can observe a gain of at least 2X the brightness over the previous approach in the relevant range of working distances.

\section{Conclusions}
In this work, we have proposed novel methods for the design and manufacturing of optimized retro reflective markers. The aim was to circumvent the material restrictions in our intended application and obtain a better response using the allowed fused silica glass material by changing the distance to and the curvature of the reflective surface.

We have identified three alternative design types, based on a custom made biconvex lens array, an off the shelf plano convex lens array and an array of glass spheres. We analysed the options from performance, manufacturing and practicality points of view and decided on the most advantageous type for our application, namely the ball lens option.

We proposed an efficient and scalable method for manufacturing the most advantageous option. We evaluated the performance of the developed marker through ray tracing simulations and in-lab experiments. The developed marker was shown to increase significantly the amount of light returned to the camera in a photogrammetric setup.

The marker is effectively easier to discriminate from the surrounding environment than the previous solution, either by automatic segmentation or manual identification by an operator. It has the potential of being a key enabling element of pose estimation in challenging visual conditions.

\section*{Acknowledgements}
The work leading to this publication has been funded by Fusion for Energy under Grant F4E-GRT-0901. This publication reflects the views only of the authors, and Fusion for Energy cannot be held responsible for any use which may be made of the information contained herein.

\bibliography{mybibfile}
\end{document}